\documentstyle[preprint,eqsecnum,aps,epsf]{revtex}

\tightenlines
\begin{document}
\preprint{\vbox{
\hbox{INPP-UVA-00-01} 
}}
\draft
\def\vp{{\bf p}}
\def\ko{K^0}
\def\kb{\bar{K^0}}
\def\al{\alpha}
\def\ab{\bar{\alpha}}
\def\be{\begin{equation}}
\def\en{\end{equation}}
\def\bea{\begin{eqnarray}}
\def\eea{\end{eqnarray}}
\def\non{\nonumber}
\def\la{\langle}
\def\ra{\rangle}
\def\epp{\epsilon^{\prime}}
\def\vep{\varepsilon}
\def\to{\rightarrow}
\def\up{\uparrow}
\def\dw{\downarrow}
\def\ms{\overline{\rm MS}}
\def\ums{{\mu}_{_{\overline{\rm MS}}}}
\def\u{\mu_{\rm fact}}

\def\pr{{\sl Phys. Rev.}~}
\def\ijmp{{\sl Int. J. Mod. Phys.}~}
\def\jp{{\sl J. Phys.}~}
\def\mpl{{\sl Mod. Phys. Lett.}~}
\def\prp{{\sl Phys. Rep.}~}
\def\prl{{\sl Phys. Rev. Lett.}~}
\def\pl{{\sl Phys. Lett.}~}
\def\np{{\sl Nucl. Phys.}~}
\def\ppnp{{\sl Prog. Part. Nucl. Phys.}~}
\def\zp{{\sl Z. Phys.}~}

\title{Substructure of the Nucleon in the Chiral Quark Model$^*$\\}

\author{X. Song$^+$}

\address{Institute of Nuclear and Particle Physics\\
Department of Physics, University of Virginia\\
Charlottesville, VA 22901, USA\\}

\maketitle
\begin{abstract}
The spin and orbital angular momentum carried by different quark 
flavors in the nucleon are calculated in the SU(3) chiral quark 
model with symmetry-breaking. The similar calculation is also 
performed for other octet and decuplet baryons. Furthermore, the 
flavor and spin contents for charm and anti-charm quarks are 
predicted in the SU(4) symmetry breaking chiral quark model.
\end{abstract}
\bigskip

\footnote{* Plenary talk presented at the Circum-Pan-Pacific RIKEN
Symposium on `High Energy Spin \\
~~ Physics', RIKEN, Wako, Japan, November 3-6, 1999.\\
$^+$\ E-mail address:\ xs3e@virginia.edu}


\leftline{\bf I. Introduction}

One of the important tasks in hadron physics is to reveal the
internal structure of the nucleon. This includes the study
of flavor, spin and orbital angular momentum shared by the 
quarks and gluons in the nucleon. These structures determine
the basic properties of the nucleon: spin, magnetic moment, 
axial coupling constant, elastic form factors, and the deep 
inelastic structure functions. The polarized deep-inelastic 
scattering (DIS) data \cite{emc,smc97} indicate 
that the quark spin only contributes about one third of the 
nucleon spin or even less. A natural and interesting question 
is: where is the {\it missing} spin ? Intuitively, and also from 
quantum chromodynamics (QCD) \cite{ji-jm}, the nucleon 
spin can be decomposed into the quark and gluon contributions
$$
{1\over 2}=<J_z>_{q+\bar q}+<J_z>_{G}={1\over 2}\Delta\Sigma+
<L_z>_{q+\bar q}+<J_z>_{G}
\eqno (1)
$$
where $\Delta\Sigma=\sum\limits_q[\Delta q+\Delta\bar q]$ and
$<L_z>_{q+\bar q}$ are the total helicity and orbital angular 
momentum carried by quarks and antiquarks. $<J_z>_{G}$ is the gluon 
angular momentum. The smallness of ${1\over 2}\Delta\Sigma$ implies 
that the missing part should be contributed either by the quark orbital
motion or the gluon angular momentum. Most recently, it has been shown 
that $<J_z>_{q+\bar q}$ might be measured in the deep virtual Compton 
scattering process \cite{ji-jws}, and one may then obtain the quark 
orbital angular momentum from the difference $<J_z>_{q+\bar q}-{1\over
2}\Delta\Sigma$. Hence the study of the quark spin and orbital angular 
momentum are important and interesting both experimentally and
theoretically.

In the naive quark model, $<L_z>_{q+\bar q}=0$ and $<L_z>_G=0$. 
In the bag model \cite{cmbag}, ${1\over 2}\Delta\Sigma \simeq 0.39$, 
and $<L_z>_{q}\simeq 0.11$, while in the Skyrme model \cite{skyrme}, 
$\Delta G=\Delta\Sigma=0$, and $<L_z>={1\over 2}$. Most recently Casu 
and Sehgal \cite{cs97} show that to fit the baryon magnetic moments and 
polarized DIS data, a large collective orbital angular momentum $<L_z>$, 
which contributes almost $80\%$ of nucleon spin, is needed. Hence the 
question of how much of the nucleon spin is coming from the quark orbital 
motion remains. 
\bigskip

\leftline{\bf II. SU(3) Chiral Quark Model}

\leftline{\qquad\bf (a) Basic Asuumptions}

The effective interaction Lagrangian for SU(3) chiral quark model 
\cite{song9802206} is 
$${\it L}_I=g_8{\bar q}\pmatrix{{G}_u^0
& {\pi}^+ & {\sqrt\epsilon}K^+\cr 
{\pi}^-& {G}_d^0
& {\sqrt\epsilon}K^0\cr
{\sqrt\epsilon}K^-& {\sqrt\epsilon}{\bar K}^0
&{G}_s^0 
\cr }q, 
\eqno (2a)$$
where ${G}_{u(d)}^0$ and ${GB}_{s}^0$ are defined as
$${G}_{u(d)}^0=+(-){\pi^0}/{\sqrt 2}+
{\sqrt{\epsilon_{\eta}}}{\eta^0}/{\sqrt 6}+
{\zeta'}{\eta'^0}/{\sqrt 3},
~~~{G}_s^0=-{\sqrt{\epsilon_{\eta}}}{\eta^0}/{\sqrt 6}+
{\zeta'}{\eta'^0}/{\sqrt 3}.
\eqno (2b)$$
The breaking effects are explicitly included. $a\equiv|g_8|^2$ denotes 
the transition probability of chiral fluctuation or splitting 
$u(d)\to d(u)+\pi^{+(-)}$, and $\epsilon a$ denotes the probability 
of $u(d)\to s+K^{-(0)}$. Similar definitions are used for 
$\epsilon_\eta a$ and $\zeta'^2a$. If the breaking is
dominated by mass suppression effect, one reasonably expects
$0\leq\zeta'^2a<\epsilon_{\eta}a\simeq\epsilon a\leq a$.

The basic {\it assumptions} of the chiral quark model are: (i) the 
quark flavor, spin and orbital contents of the nucleon are determined 
by its valence quark structure and all possible chiral fluctuations, 
and probabilities of these fluctuations depend on the interaction 
Lagrangian (2), (ii) the coupling between the quark and Goldstone 
boson is rather weak, one can treat the fluctuation $q\to q'+{\rm GB}$ 
as a small perturbation ($a\sim 0.10-0.15$) and the contributions from 
the higher order fluctuations can be neglected, and (iii) quark spin-flip
interaction dominates the splitting process $q\to q'+{\rm GB}$. This 
can be related to the picture given by the instanton model \cite{ins}, 
hence the spin-nonflip interaction is suppressed.

Based upon the assumptions, the quark {\it flips} its spin and 
changes (or maintains) its flavor by emitting a charged (or neutral) 
Goldstone boson. The light quark sea asymmetry $\bar u<\bar d$ is 
attributed to the existing {\it flavor asymmetry} of the valence quark 
numbers (two valence $u$-quarks and one valence $d$-quark) in the 
proton. On the other hand, the quark spin reduction is due to the 
{\it spin dilution} in the chiral splitting processes. Furthermore,
the quark spin component changes one unit of angular momentum,
$(s_z)_f-(s_z)_i=+1$ or $-1$, due to spin-flip in the fluctuation 
with GB emission. The angular momentum conservation requires the 
{\it same} amount change of the orbital angular momentum but with 
{\it opposite sign}, i.e. $(L_z)_f-(L_z)_i=-1$ or $+1$. This {\it 
induced} orbital motion is distributed among the quarks and antiquarks, 
and compensates the spin reduction in the chiral splitting. 

\leftline{\quad \bf~(b) Quark Spin Contents in the Nucleon}

The spin-weighted quark contents are
$$\Delta u^p={4\over 5}\Delta_3-a,~~\Delta d^p=-{1\over 5}\Delta_3-a,~~
\Delta s^p=-\epsilon a,
\eqno (3a)
$$
where $\Delta_3={5\over 3}[1-a(\epsilon+2f)]$ and 
$f\equiv{1\over 2}+{{\epsilon_{\eta}}\over 6}+{{\zeta'^2}\over 3}$. The 
total quark spin content in the proton is
$${1\over 2}\Delta\Sigma^p={1\over 2}(\Delta u^p+\Delta d^p+\Delta
s^p)={1\over 2}-a(1+\epsilon+f)\equiv {1\over 2}-a\xi_1 
\eqno (3b)
$$
where the notation $\xi_1\equiv 1+\epsilon+f$ is defined. 
A special feature of the chiral quark model is that {\it all the
spin-weighted antiquark contents are zero}
$$\Delta\bar q=0.
\eqno (3c)
$$
Hence $(\Delta q)_{sea}\neq \Delta\bar q$, which differs from
the predictions of the sea quark and antiquark pair produced by
a gluon (see discussion in \cite{smw}). 

\leftline{\quad (c)~\bf Quark Orbital Momentum in the Nucleon}
 
The orbital angular momentum produced in the splitting $q_{\up}\to q'_{\dw}
+{\rm GB}$ is shared by the recoil quark ($q'$) and the Goldstone boson
(GB). Defining $2\kappa$ as the fraction of the orbital angular momentum
shared by the GB, then the fraction shared by the recoil quark is 
$1-2\kappa$. We assume the fraction of $2\kappa$ is equally shared by 
the quark and antiquark in the GB and call $\kappa$ the {\it partition 
factor} which satisfies $0<\kappa<1/2$. For $\kappa=1/3$, the
three particles $-$ the recoil quark, quark and antiquark in the
GB equally share the induced orbital angular momentum. For the proton, 
we obtain
$$<L_z>_{q}^p\equiv<L_z>_{u+d+s}^p=(1-\kappa)\xi_1a
\eqno (4a)$$
$$<L_z>_{\bar q}^p\equiv
<L_z>_{\bar u+\bar d+\bar s}^p=\kappa\xi_1a
\eqno (4b)$$
and
$$<L_z>_{q+\bar q}^p\equiv<L_z>_{q}^p+<L_z>_{\bar q}^p=\xi_1a
\eqno (4c)$$
The orbital angular momentum of each quark flavor may depend on the 
partition factor $\kappa$, but the total orbital angular momentum (4c) 
is independent of $\kappa$. Furthermore, the amount $\xi_1a$ is just 
the same as the total spin reduction in (3b), and the sum of (4c) and 
(3b) gives
$$<J_z>_{q+\bar q}^p={1\over 2}\Delta\Sigma_{q+\bar q}^p+<L_z>_{q+\bar
q}^p={1\over 2}
\eqno (4d)$$
In the chiral fluctuations, the missing part of the quark spin is 
{\it transferred} into the orbital motion of quarks and antiquarks. 
The amount of quark spin reduction $a(1+\epsilon+f)$ in (3b) is 
canceled by the equal amount increase of the quark orbital angular
momentum in (4c), and the total angular momentum of nucleon is unchanged. 

\leftline{\quad (d)~\bf Paramters}

The model parameters are determined by three inputs, $\Delta u-\Delta
d=1.26$, $\Delta u+\Delta d-2\Delta s=0.60$, and $\bar d-\bar u=0.143$.
The result is: $a=0.145$, $\epsilon=0.46$, and $\zeta'^2=0.10$. The 
orbital angular momenta shared by different quark flavors are listed in
Table I. We plot the orbital angular momenta carried by quarks and 
antiquarks in the proton as function of $\kappa$ in Fig.1. 
Using the parameter set given above, $<L_z>_{q+\bar q}^p
\simeq 0.30$, i.e. nearly $60\%$ of the proton spin is coming from the 
orbital motion of quarks and antiquarks, and $40\%$ is contributed by 
the quark and antiquark spins. Comparison of our result with other models 
is given in Fig.2. 

Extension to other baryons and application to the baryon magnetic moments 
were discussed in \cite{song9802206}. It has been shown that although the 
chiral model result of the magnetic moments seems to be better than the
nonrelativistic quark model result, there is no significant difference 
between them. This is because the positive orbital contribution to the 
magnetic moment cancels in part the negative contribution given by the
quark spin reduction. This cancellation was also discussed in \cite{cl3}. 
Hence the magnetic moment might not be a good observable to manifest the
quark orbital contribution.
 
\leftline{\bf II. SU(4) Chiral Quark Model}

The effective intercation Lagrangian in SU(4) case is
$${\it L}_I=g_{15}{\bar q}\pmatrix{{G}_u^0
& {\pi}^+ & {\sqrt{\epsilon}}K^+ & {\sqrt{\epsilon_c}}{\bar D}^0 \cr 
{\pi}^-& {G}_d^0& {\sqrt{\epsilon}}K^0 &{\sqrt{\epsilon_c}}D^-\cr
{\sqrt{\epsilon}}K^-& {\sqrt{\epsilon}}{\bar K}^0&{G}_s^0 &
{\sqrt{\epsilon_c}}{D}_s^-\cr 
{\sqrt{\epsilon_c}}{D}^0 & {\sqrt{\epsilon_c}}{D}^+
& {\sqrt{\epsilon_c}}{D}_s^+ &{G}_c^0 \cr }q, 
\eqno (6)$$
where ${G}_{u(d)}^0$ and ${G}_{s}^0$ are defined similarly as in (2b), 
but with additional $\epsilon_c$ term, and 
${G}_c^0=-{\zeta'}{{{\sqrt 3}\eta'^0}\over 4}+
{\sqrt{\epsilon_c}}{{3\eta_c}\over 4}$, with $\eta_c=(c\bar c)$.

In the SU(4) chiral quark model, the charm and anticharm quarks are
produced nonperturbatively, and they are `{\it intrinsic}. The intrinsic
charm helicity $\Delta c$ is nonzero and definitely negative. 
To estimate the size of $\Delta c$ and other intrinsic charm
contributions, we use the same parameter set ($a=0.145$,
$\epsilon\simeq\epsilon_\eta=0.46$, $\zeta'^2=0.10$) given in the SU(3) 
case, and leave $\epsilon_c$ as a variable, then other quark
flavor and helicity contents can be expressed as functions of 
$\epsilon_c$. We found that $\epsilon_c\simeq 0.1-0.3$
satisfactorily describes the data. Our model results, data, and
theoretical predictions from other approaches are listed in Table II 
and Table III respectively. Several remarks are in order:
(1) our result, ${2\bar c}/\sum(q+\bar q)\simeq 3.7\%$, agrees with 
that given in \cite{bs91} and the earlier number given in \cite{dg77}. 
But the result given in \cite{hk94} is much smaller (0.5$\%$) than ours. 
(2) our prediction $\Delta c=-0.029\pm 0.015$ is very close to the 
result $\Delta c=-0.020\pm 0.005$ given in the instanton QCD vacuum model
\cite{amt98}. However the size of $\Delta c$ given in \cite{pst98} is
about two order of magnitude smaller than ours. (3) We plot the ratio
$\Delta c/\Delta\Sigma$ as function of $\epsilon_c$ in Fig.3. Our result
$\Delta c/\Delta\Sigma\simeq 0.084\pm 0.046$ agrees well with the 
prediction given in \cite{blos98} and is also not inconsistent with 
the result given in \cite{amt98}.
 
To summarize, the chiral quark model with $a$ $few$ parameters can well 
explain many existing data of the nucleon properties: (1) strong flavor
asymmetry of light antiquark sea: $\bar d> \bar u$, (2) nonzero strange 
quark content, $<\bar ss>\neq 0$, (3) sum of quark spins is small, 
$<s_z>_{q+\bar q}\simeq 0.1-0.2$, (4) sea antiquarks are not polarized: 
$\Delta\bar q\simeq 0$ ($q=u,d,...$), (5) polarizations of the sea 
quarks are nonzero and negative, $\Delta q_{sea}< 0$, (6) the 
orbital angular momentum of the sea quark is parallel to the proton 
spin, and (7) the SU(4) chiral quark model predicts a small amount 
of intrinsic charm and a negative $\Delta c$ in the proton. (1)-(4) 
are consistent with data, and (5)-(7) could be tested by future 
experiments. 
\bigskip

\leftline{\bf Acknowledgments}

The author would like to thank S. Brodsky, P. K. Kabir and H. J. Weber 
for useful comments and suggestions. This work was supported in part by 
the U.S. DOE Grant No. DE-FG02-96ER-40950, the Institute of Nuclear and 
Particle Physics, University of Virginia, and the Commonwealth of
Virginia.

\bigskip

\begin{table}[ht]
\begin{center}
\caption{Quark spin and orbital angular momentum in the proton in
different models.}
\begin{tabular}{ccccccc} 
Quantity & Data \cite{smc97}   && This paper& &  CS \cite{cs97} & NQM\\ 
&& $\kappa=1/4$ & $\kappa=1/3$& $\kappa=3/8$ &&\\
\hline 
$<L_z>_u^p$       & $-$ & 0.115   & 0.130   & 0.138  & $-$    & 0  \\ 
$<L_z>_d^p$       & $-$ & 0.073   & 0.043   & 0.027  & $-$    & 0  \\ 
$<L_z>_s^p$       & $-$ & 0.038   & 0.028   & 0.023  & $-$    & 0  \\ 
$<L_z>_{\bar u}^p$& $-$ &$-$0.003 & $-$0.003&$-$0.003 & $-$    & 0  \\ 
$<L_z>_{\bar d}^p$& $-$ & 0.057   & 0.076   & 0.085  & $-$    & 0  \\ 
$<L_z>_{\bar s}^p$& $-$ & 0.021   & 0.028   & 0.031  & $-$    & 0  \\ 
\hline
$<L_z>_{q+\bar q}^p$ & $-$ & 0.30 & 0.30    & 0.30 &  0.39   & 0  \\ 
\hline
$\Delta u^p$ & $0.85\pm 0.05$ & & 0.86     &  & 0.78    & 4/3  \\ 
$\Delta d^p$ & $-0.41\pm 0.05$ & & $-0.40$ &  & $-0.34$ & $-1/3$\\ 
$\Delta s^p$ & $-0.07\pm 0.05$ & &$-0.07$ &  & $-0.14$ & 0  \\ 
\hline
${1\over 2}\Delta\Sigma^p$ & $0.19\pm 0.06$ & & 0.20 &  & 0.08 & 1/2\\ 
\hline
\end{tabular}
\end{center}
\end{table}

\begin{table}[ht]
\begin{center}
\caption{Quark flavor observables}
\bigskip
\begin{tabular}{|c|c|c|c|} \hline
Quantity & Data   & SU(3) &  SU(4)\\
\hline 
$\bar d-\bar u$ & $0.147\pm 0.039$ & 0.147 & 0.120  \\
                & $0.110\pm 0.018$ &       &        \\
\hline 
${{\bar u}/{\bar d}}$ &$[{{\bar u(x)}\over {\bar d(x)}}]_{0.1<x<0.2}
=0.67\pm 0.06$ & 0.65 & 0.69\\
&$[{{\bar u(x)}\over {\bar d(x)}}]_{x=0.18}=0.51\pm 0.06$ & & \\ 
\hline
${{2\bar s}/{(\bar u+\bar d)}}$ & ${{<2x\bar s(x)>}\over {<x(\bar
u(x)+\bar d(x))>}}=0.477\pm 0.051$& 0.69 & 0.69\\
${{2\bar c}/{(\bar u+\bar d)}}$ & $-$ & 0 & $0.28\pm 0.14$ \\
\hline
${{2\bar s}/{(u+d)}}$ & ${{<2x\bar s(x)>}\over
{<x(u(x)+d(x))>}}=0.099\pm 0.009$& 0.128 &0.120\\
${{2\bar c}/{(u+d)}}$ & $-$ & 0 &$0.05\pm 0.02$\\
\hline
$f_s\equiv{2\bar s}/\sum(q+\bar q)$ & $0.10\pm 0.06$ & 0.10&0.09\\
       & $0.15\pm 0.03$ &     &   \\
       & ${{<2x\bar s(x)>}\over {\sum<x(q(x)+\bar q(x))>}}
=0.076\pm 0.022$ & &      \\
$f_c\equiv{2\bar c}/\sum(q+\bar q)$ & 0.03~\cite{bs91} & 0 &$0.037\pm
0.015$\\
& 0.02~\cite{dg77} &  & \\
& 0.005~\cite{hk94}&  & \\
\hline
${{\sum\bar q}/{\sum q}}$ & ${{\sum<x\bar
q(x)>}\over {\sum<xq(x)>}}=0.245\pm 0.005$ & 0.235 & 0.246\\
 $f_3/f_8$ & $0.23\pm 0.05$  & 0.21 &0.22\\
\hline
\end{tabular}
\bigskip
\end{center}
\end{table}

\begin{table}[ht]
\begin{center}
\caption{Quark spin observables}
\bigskip
\begin{tabular}{|c|c|c|c|}\hline
Quantity & Data   & SU(3) &  SU(4)\\
\hline 
$\Delta u$ & $0.85\pm 0.04$ & 0.86 &0.83\\
$\Delta d$&$-0.41\pm$0.04 &$-$0.40&$-$0.39\\
$\Delta s$&$-0.07\pm$0.04 &$-$0.07&$-$0.07\\
$\Delta c$ & $-0.020\pm 0.004$~\cite{amt98} & 0& $-0.029\pm 0.015$ \\
           & $-5\cdot 10^{-4}$~\cite{pst98} &  &          \\
\hline
$\Delta\bar u$, $\Delta\bar d$ & $-0.02\pm 0.11$&0&0\\
$\Delta\bar s$, $\Delta\bar c$ & $-$&0&0\\
\hline
$\Delta c/\Delta\Sigma$ & $-0.08\pm 0.01$~\cite{blos98} & 0& $-0.084\pm
0.046$ \\
  & $-0.033$ ~\cite{amt98}& &  \\
$\Delta c~/~c$ & $-$ &  $-$& $-$0.314  \\
\hline 
$\Gamma_1^p$ & $0.136\pm 0.016$ & 0.133 &0.133\\
$\Gamma_1^n$ & $-0.036\pm 0.007$ & $-0.037$ & $-0.034$  \\
\hline 
$\Delta_3$ &1.2573$\pm$0.0028 &1.26&1.259\\
$\Delta_8$& 0.579$\pm$ 0.025& 0.60&0.578 \\
\hline
\end{tabular}
\bigskip
\end{center}
\end{table}

\begin{figure}[h]
\epsfxsize=2.8in
\centerline{\epsfbox{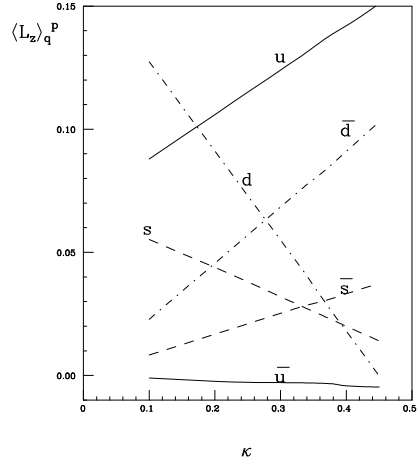}}
\caption{Quark or antiquark orbital angular momentum $<L_z>_{q,\bar q}$
in the proton as function of $\kappa$.}
\end{figure}

\begin{figure}[h]
\epsfxsize=2.8in
\centerline{\epsfbox{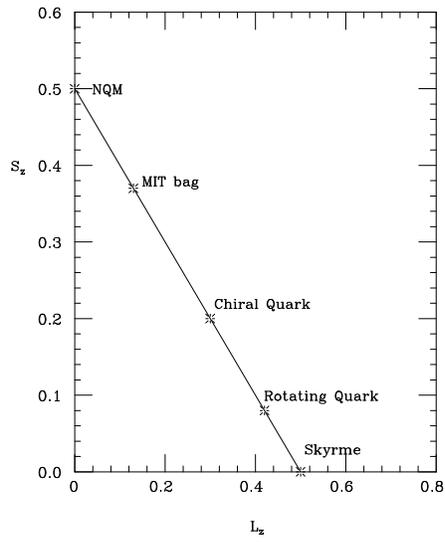}}
\caption{Quark spin and orbital angular momentum in the nucleon in 
different models.}
\end{figure}

\begin{figure}[h]
\epsfxsize=2.8in
\centerline{\epsfbox{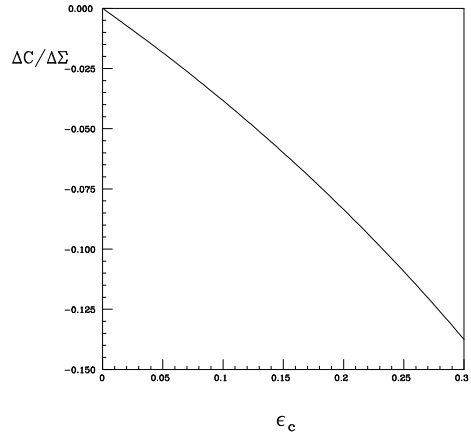}}
\caption{Intrinsic charm quark polarization in the proton as function of
$\epsilon_c$.}
\end{figure}

\end{document}